\documentclass[preprint]{aastex631}
	\shorttitle{use GC model explaning the muon excess}
	\shortauthors{Liu, Yang \& Ruan}
	\usepackage{amsmath}
	
	\graphicspath{{./}{figures/}}
\begin{document}
	
	\title{Explaining muon excess in cosmic rays using the gluon condensation model}
	
	\author{Bingyang Liu}
	\affil{Department of Physics, East China Normal University, 500 Dongchuan Road, Shanghai 200241, China}
	
	\author{Zhixiang Yang}
	\affil{Department of Physics, East China Normal University, 500 Dongchuan Road, Shanghai 200241, China}
	
	\author{Jianhong Ruan}
	\affil{Department of Physics, East China Normal University, 500 Dongchuan Road, Shanghai 200241, China}
	
	\correspondingauthor{JianHong Ruan}
	\email{jhruan@phy.ecnu.edu.cn}

\begin{abstract}
		Ultrahigh-energy cosmic rays are often characterized indirectly by analyzing the properties of secondary cosmic ray particles produced in the  collisions with air nuclei. The particle number $N_\mu$ of muon and the depth of shower maximum $X_\mathrm{max}$ after air shower cascade are mostly studied to infer the energy and mass of the incident cosmic rays. 
		Research have shown that there is a significant excess in the observed number of muons arriving at the ground from extensive air showers (EAS) compared to the simulations using the existing cosmic ray hadronic interaction model. To explain this muon excess phenomenon, a new theoretical model, the gluon condensation model (GC model), is introduced in this paper and simulated by using the AIRES engine.
		We assume that the GC effect appears mainly in the first collision of the cascade leading to a significant increase in the strangeness production, consequently, the production rate of kaons is increased and $n_K/n_\pi$ is greater than the value of the usual hadronic interaction process. 
		In the calculation, the model assumes that only pions and kaons are produced in GC state. The  increase of strange particle yield would mean that the energy transferred from the hadronic cascade to electromagnetic cascade through $\pi^{0} \rightarrow 2\gamma$ decay is reduced. This would in turn increase the number of muons at the ground level due to meson decays.
		Our model provides a new theoretical possibility to explain the muon excess puzzle.
	\end{abstract}
	
	
\section{Introduction} 
\label{sec:Introduction}
	
When high-energy cosmic rays enter the Earth's atmosphere from outer space, they interact with atmospheric nuclei (primarily nitrogen, oxygen, and argon), leading to particle cascades and the production of a series of secondary particles such as pions, kaons, and baryons. In the process of air shower cascades, the production of muons is closely related to the decay of these secondary particles, with about $90\%$ originating from the decay of pions and kaons. Muons propagate with minimal energy loss in the atmosphere, nearly undecayed, and eventually reach the Earth's surface. Based on the highest-energy cosmic ray data observed so far, the Pierre Auger Observatory \cite[]{PierreAuger:2014ucz,aab2016testing,aab2016pierre} and Telescope Array\cite[]{aab2017combined} indicated a significant discrepancy in the muon number within extensive air showers (EAS) compared to the data predicted by the hadronic interaction models using in detectors like LHC \cite[]{PhysRevD.94.032007} and SPS \cite[]{NA61SHINE:2022tiz}. Hadronic interaction models such as Sibyll-2.1 \cite[]{ahn2009cosmic}, QGSJetII-04 \cite[]{ostapchenko2011monte}, and EPOS-LHC \cite[]{pierog2015epos} are proposed after further development and adjustment to the data. However, the observed number of muons still exceed the results simulated with these models, termed as the muon excess. 		

The number of muons $N_\mu$ and depth of shower maximum $X_\mathrm{max}$ are the key parameters for studying the cosmic rays, which are sensitive to several properties of hadronic interactions. The mean $N_\mu$ is mainly sensitive to the multiplicity, the energy fraction of $\pi^0$ (which carries the proportion of the incident energy in hadronic interactions and then transfers to electromagnetic interactions), and the primary mass of cosmic rays. The mean $X_{\mathrm{max}}$ is mainly sensitive to the cross-section, elasticity, multiplicity, and primary mass \cite[]{allen2013testing}, and the measurement of $X_\mathrm{max}$ shows well agreement with the model estimates.
It can be seen that only by altering the energy fraction of $\pi^0$, $N_\mu$ can be increased while $X_{\mathrm{max}}$ is kept unchanged. To address the muon excess issue, the "fireball model" proposed by \cite{anchordoqui2017strange} enhances the strangeness to suppress $\pi^0$ production. Additionally, in \cite{anchordoqui2022explanation}, a model is proposed where swapping $\pi$ and $K$ particles is used to explain the muon excess, with $f_s$ representing the swapping probability of $\pi$ to $K$. The work of \cite{baur2023core} employs a simplified core-corona method to estimate the muon production when statistical hadronization is present. The core part corresponds to the normal particles produced by hadronic interactions, while the corona part corresponds to the production of "statistical" or "thermal" particles, each with its own weight summing up to 1.  
Moreover, a novel study\cite[]{PierreAuger:2021qsd} on the shower-to-shower fluctuations of $N_\mu$ by Pierre Auger Observatory suggests that this muon deficit in models might be a small deficit at every stage of the shower that accumulates along the shower development, and they also mentioned that a very particular modification of the first interaction may also change $N_\mu$ without changing the fluctuations. Although the research about muon puzzle are becoming increasingly detailed, the truth remains elusive, necessitating the further investigations.

Regarding the muon excess issue, we tend to believe that it's the initial collision in air showers which cause the main difference between the measurements and model predictions. In ultrahigh-energy air showers, the interaction energy of their first collision may significantly exceed the energy of accelerators, potentially leading to new physical phenomena. Zhu et al., \cite[]{zhu1999new,zhu2004properties} have proposed that in ultrahigh-energy interactions, the gluon distribution in hadrons may undergo condensation, which would influence the hadronization products and subsequently the final muon production. The aim of this paper is to use the gluon condensation model in the first collision of the cascades to resolve the muon excess problem.

Air shower cascades are driven by collisions between ultrahigh energy primary cosmic rays with the hadrons in atmosphere. During the collisions, the parton distribution in small $x$ region is dominated by gluons, where $x$ is the parton momentum fraction (Bjorken-$x$). The evolution of gluons follows the equations of quantum chromodynamics (QCD), but in the small $x$ region, traditional evolution equations should be modified due to the recombination among partons. The modified equation, known as the Zhu-Shen-Ruan (ZSR) equation \cite[]{zhu1999new,zhu2004properties}, predict that gluons in hadrons may converge at a critical momentum $k_c$ in very high energy region, which is called  the  gluon condensation (GC). By simulation and analysis of the gluon condensation effects in collision processes, we found that the GC effect significantly increases the production of pions and kaons, especially the production rate of $n_K/n_\pi$  being much higher than that in normal collidings \cite[]{wong1994introduction}. The increase of kaon yield would mean that the energy transferred from the hadronic cascade to electromagnetic cascade through $\pi^{0} \rightarrow 2\gamma$ decay is reduced. This would in turn increase the number of muons at the ground level due to meson decays, which likely resolve the problem of muon excess.

The paper is structured as follows: In section \ref{sec:2} the basic principles of the gluon condensation model are introduced. Section \ref{sec:3} provides a detailed explanation of our work in addressing the muon excess puzzle, including the mechanism of muon production, the production of pions and kaons under different conditions of high-energy collisions, and the key parameters and data analysis of the simulation process in GC conditions. Finally, we present discussions and conclusions in \autoref{sec:4}.

\section{Introduction to gluon condensation model}
\label{sec:2}
\subsection{QCD dynamic evolution equation}
\begin{figure}%[H] 
	\centering
	\includegraphics[width=0.6\textwidth]{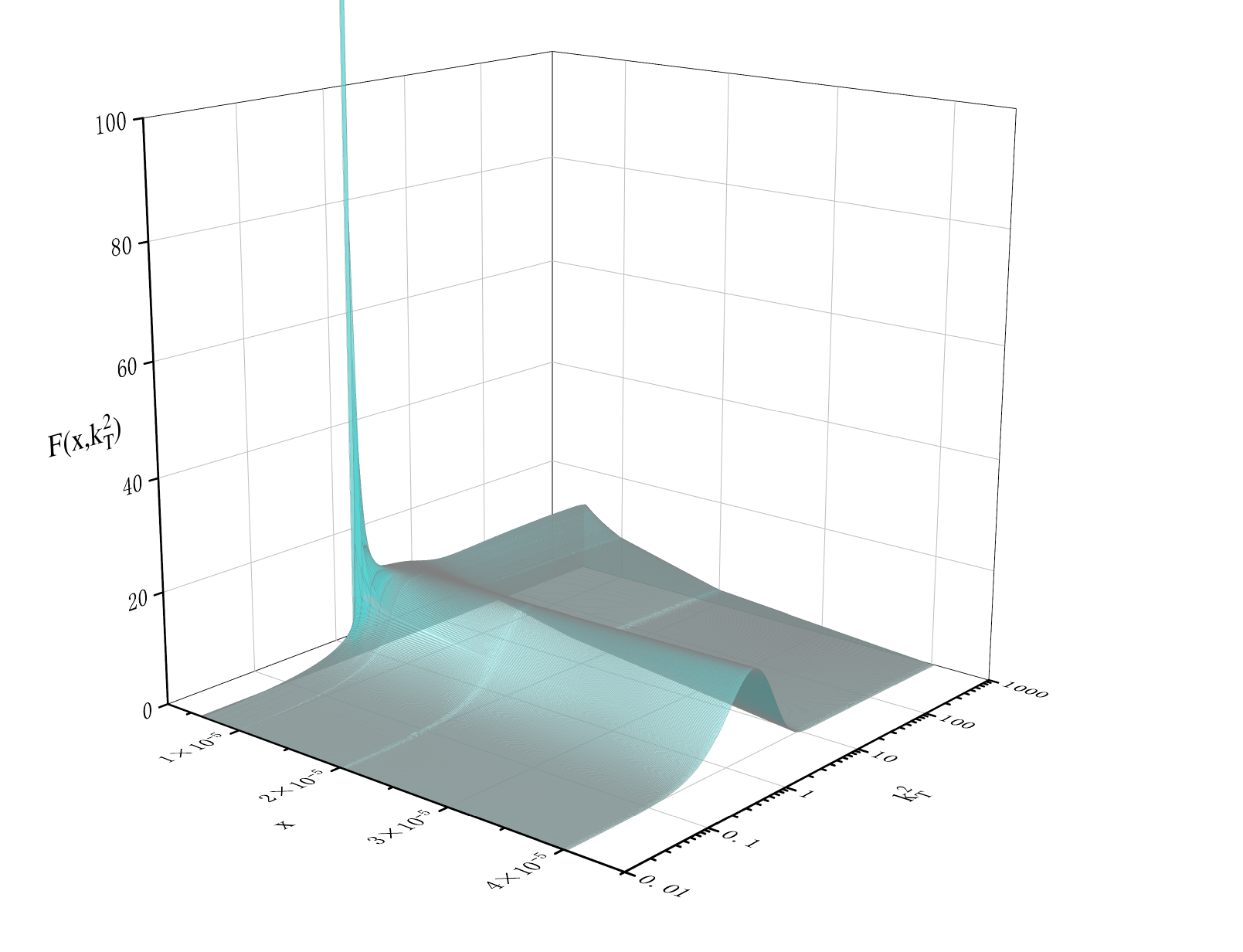}
	\caption{GC solution of the ZSR equation}	
	\label{fig.1}
\end{figure}

In the infinite frame of momentum, hadrons are composed of  partons, namely quarks and gluons. Partons evolve according to the evolution equations based on Quantum Chromodynamics (QCD) within the framework of the standard model. DGLAP (Dokshitzer-Gribov-Lipatov-Altarelli-Parisi) equation\cite[]{gribov1972deep,dokshitzer1977calculation,altarelli1977asymptotic} and BFKL (Balitsky-Fadin-Kuraev-Lipatov) equation\cite[]{fadin1975pomeranchuk,kuraev1976multireggeon,balitsky1978pomeranchuk} are the standard parton evolution equations. In small $x$ region, BFKL equation is more suitable since the parton evolution in it includes the information of transverse momentum, and the transverse momentum is one of the key parameters when considering the parton recombinations.  
The BFKL equation reads\cite[]{Enberg:2005zj}

\begin{equation} 
	\label{eq.1}
	-x\frac{\partial F(x,k_{\perp}^{2})}{\partial x} =\frac{3\alpha_{s}k_{\perp}^{2}}{\pi}\int_{k_{0}^{2}}^{\infty}\frac{dk_{\perp}^{\prime2}}{k_{\perp}^{\prime2}}\left\{\frac{F(x,k_{\perp}^{\prime2})-F(x,k_{\perp}^{2})}{|k_{\perp}^{\prime2}-k_{\perp}^{2}|}+\frac{F(x,k_{\perp}^{2})}{\sqrt{k_{\perp}^{4}+4k_{\perp}^{\prime4}}}\right\},
\end{equation}
where the function $F(x,k_{\perp}^{2})$ is the unintegrated gluon distribution function and $k_{\perp}$ is its transverse momentum, $\alpha_s$ is the effective QCD coupling constant. The BFKL equation describes the parton evolution process of one parton (with larger $x$) splitting into two partons (with smaller $x$). So, as $x$ decreases, the parton density increases, and gradually, the interactions between partons become non-negligible. The ZSR equation is a modified-BFKL equation which considered the parton recombination effects of twist-4 processes using the Time-Ordered Perturbative Theory(TOPT) \cite[]{zhu1999new,zhu2004properties}.
The ZSR equation is  

\begin{equation}
	\begin{aligned}
		&-x\frac{\partial F(x,k_{\perp}^{2})}{\partial x} \\
		&=\frac{3\alpha_{s}k_{\perp}^{2}}{\pi}\int_{k_{0}^{2}}^{\infty}\frac{dk_{\perp}^{\prime2}}{k_{\perp}^{\prime2}}\left\{\frac{F(x,k_{\perp}^{\prime2})-F(x,k_{\perp}^{2})}{|k_{\perp}^{\prime2}-k_{\perp}^{2}|}+\frac{F(x,k_{\perp}^{2})}{\sqrt{k_{\perp}^{4}+4k_{\perp}^{\prime4}}}\right\} \\
		&-\frac{81}{16}\frac{\alpha_{s}^{2}}{\pi R_{N}^{2}}\int_{k_{0}^{2}}^{\infty}\frac{dk_{\perp}^{\prime{2}}}{k_{\perp}^{\prime{2}}}\left\{\frac{k_{\perp}^{2}F^{2}(x,k_{\perp}^{\prime{2}})-k_{\perp}^{\prime{2}}F^{2}(x,k_{\perp}^{2})}{k_{\perp}^{\prime{2}}|k_{\perp}^{\prime{2}}-k_{\perp}^{2}|}+\frac{F^{2}(x,k_{\perp}^{2})}{\sqrt{k_{\perp}^{4}+4k_{\perp}^{\prime{4}}} }\right\} .
	\end{aligned}
	\label{eq.2}
\end{equation}	
The last term in the above equation is the contribution of gluon recombination twist-4 Feynman diagrams, where $R_N$ is the effective radius of nucleon. This specific nonlinear term may couple stochastic perturbations, introducing complexity and instability into the system, magnifying small changes in initial conditions, and leading to chaotic phenomena. One criterion for the appearance of chaotic phenomena is the presence of a positive Lyapunov exponent \cite[]{wei2008can,zhu2016chaotic,zhu2017gluon}, indicating the system high sensitivity to small changes in initial conditions. However, the chaos alone cannot produce the GC solutions in the ZSR equation. The TOPT regularized nonlinear kernel plays its second important role for converting the chaotic vibration, which occurs in the evolution equation when the Bjorken variable $x$ is less than the critical point $x_c$, into the strong shadowing and antishadowing effect. The GC effect results from the combined influence of chaos and antishadowing effect. Potential perturbation lead to chaotic oscillations in the BFKL dynamics, generating the positive antishadowing corrections, which then convert the shadowed gluons to the gluons with the critical momentum $\left(x_c, k_c\right)$ \cite[]{zhu2022gluon}. The gluon condensation is a natural consequence of gluons converging to the critical momentum state. The specific distribution function of gluons can be approximately represented as a delta function, as illustrated in \autoref{fig.1}.

High energy collisions between hadrons can be regarded as the interactions between their internal quarks and gluons, with gluons playing  dominant roles in small $x$ region, where the ZSR equation should be used. In QCD, the yield of secondary particles in high energy proton-proton collisions are related to the number of the excited gluons participating in multiple interactions. The energy of  proton-proton collisions in ultrahigh energy cosmic ray may exceed the energy threshold required to generate the gluon condensation.

\subsection{Particle generation in GC state}

To deeply understand the hadronization process in QCD, a recent research by Roberts \cite[]{roberts2020insights} based on the Dyson-Schwinger (DS) equations showed that the effective quark mass in hadrons might be dynamically generated due to the gluon nonlinearity \cite[]{mathieu2010pos,chang2014basic,rodriguez2020process}. In this mechanism the nonperturbative gluon propagator is constructed by lots of partons (i.e., gluons and sea quarks) in the hadron infinite momentum frame. Massless quark gains a small current mass through the Higgs mechanism and then further acquires the effective mass through the nonperturbative gluon propagator. In high-energy proton-proton collisions, the produced asymptotically free quarks and gluons cannot exist independently due to color confinement. These quarks and gluons interact by exchanging gluons, forming stable bound states known as hadrons. The gluon condensates, which are rich in gluons, greatly generate the number of the constituent quarks, thereby making the hadronization process more effective and increasing the number of pions and kaons. As a limit, it is imaged that at the center-of-mass system almost all available collision energy is used to produce pions and kaons. In this paper, the gluon condensation model is exploited to calculate the number of secondary particles of proton-proton collisions in the high-energy cosmic ray. The AIRES simulation engine \cite[]{sciutto1999aires}, which provides full space-time particle propagation in a realistic environment, is adopted to describe the shower evolution. 

The reaction process of hadronic collisions can be given by

\begin{equation} 
	p+p\rightarrow p+p+\pi+K 
	\label{eq.3}
\end{equation}

At the center-of-mass system, considering the relativistic invariance and energy conservation, one straightly writes

\begin{equation} 
	\begin{aligned}
		&\left(2m_p^2+2E_pm_p\right)^{\frac{1}{2}}=E_{p1}^*+E_{p2}^*+n_\pi m_\pi+n_km_k\\
		&E_p+m_p=m_p\gamma_1+m_p\gamma_2+n_\pi m_\pi\gamma_3+n_km_k\gamma_4
	\end{aligned}
	\label{eq.4}
\end{equation}
where $E_{pi}^*$ is the energy of the leading proton at the center-of-mass system, $\gamma_i$ is the corresponding Lorentz factor, $n_\pi$ and $n_K$ are the number of pions and kaons produced after the collision, respectively. In this work, we assume that the secondary particles resulting from the collisions between the incident protons with the proton targets, specifically including leading protons, pions, and kaons. These secondary particles (protons, pions, and kaons) continue to collide with protons in the atmosphere until they reach a critical energy, where particle decay becomes the main process. Using the inelasticity $K \sim 0.5$ \cite[]{gaisser2016cosmic}, we get

\begin{equation}
	\begin{aligned}
		&E_{p1}^*+E_{p2}^*=\left(\frac{1}{K}-1\right)\left(n_\pi m_\pi+n_km_k\right)\\
		&m_p\gamma_1+m_p\gamma_2=\left(\frac{1}{K}-1\right)\left(n_\pi m_\pi\gamma_3+n_km_k\gamma_4\right).
	\end{aligned}
	\label{eq.5}
\end{equation}

Assuming that the Lorentz factors for pions and kaons are the same, then $E_\pi/E_K = m_\pi/m_K$. At the GeV scale, setting $n_K = c n_\pi$, one can get

\begin{equation} 
	\begin{aligned}
		&\ln n_\pi=0.5\ln(m_{p}+E_{p})+a^{\prime} \\
		&\ln n_\pi=\ln(cE_{k}+E_{\pi})+b^{\prime}, 
	\end{aligned}
	\label{eq.6}
\end{equation}
where
\begin{equation} 
	\begin{aligned}
		&a^{\prime}=0.5\ln2m_{p}-\ln(cm_{k}+m_{\pi})+\ln\frac{1}{2} \\
		&b^{\prime}=\ln2m_{p}-2\ln(cm_k+m_\pi)+\ln\frac{1}{2}.
	\end{aligned}
	\label{eq.7}
\end{equation}

Taking Eqs.\ref{eq.3}, \ref{eq.4}, \ref{eq.5} as the basis of derivation, when the primary energy of the incident proton or heavy nuclei is known, we can combine Eq.5 and Eq.6 to calculate the relationship between the number of $\pi$, $K$ produced after the first collision. Then, the number and energy of pions and kaons will serve as the initial data in the AIRES simulation in the next step.

\section{Explanation of the muon excess phenomenon}
\label{sec:3}
\subsection{Mechanism of muon production}
\label{sec:3.1}		
\begin{figure}[htbp]
	\centering 
	\includegraphics[width=0.42\textwidth]{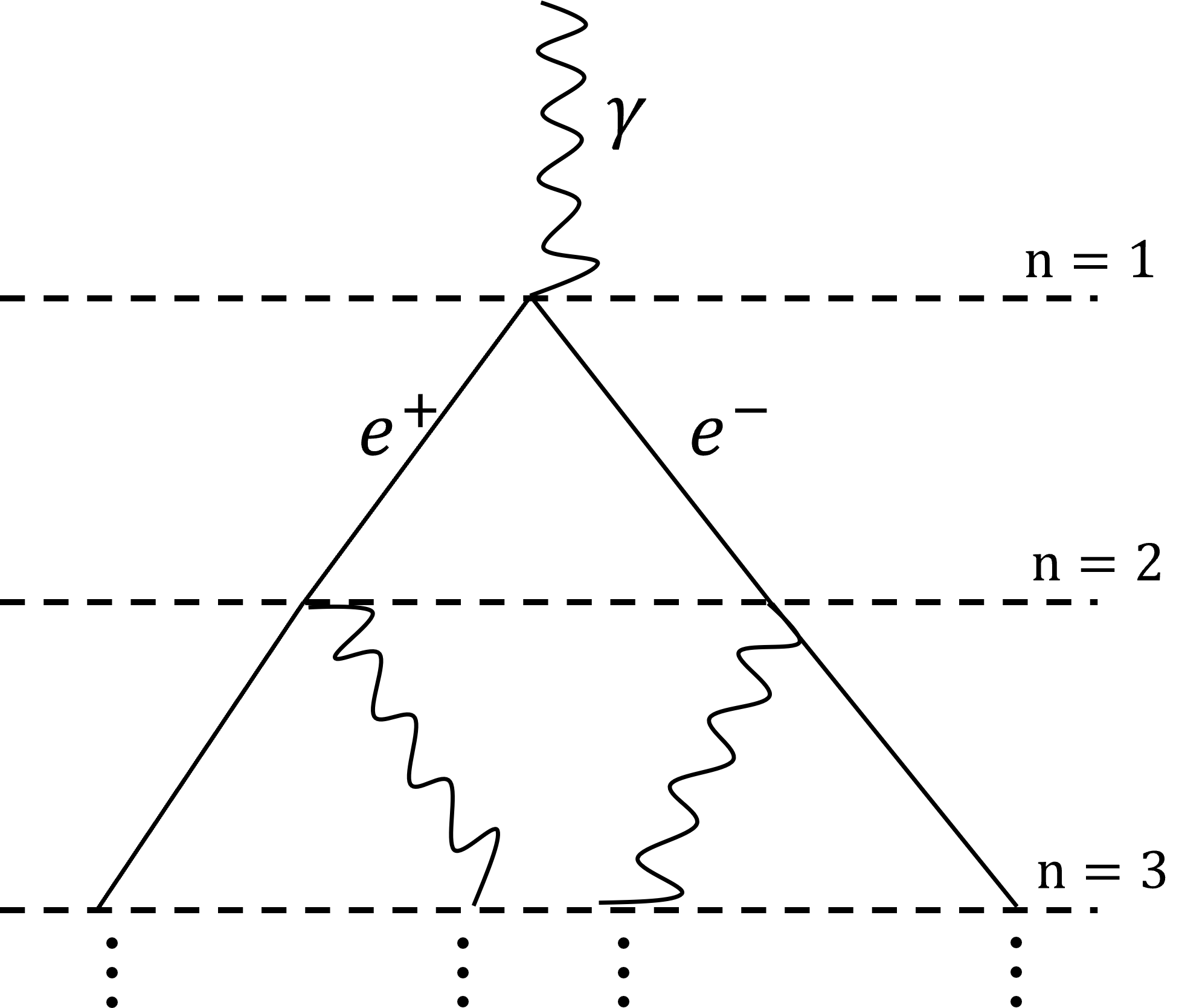}\hfill
	\includegraphics[width=0.42\textwidth]{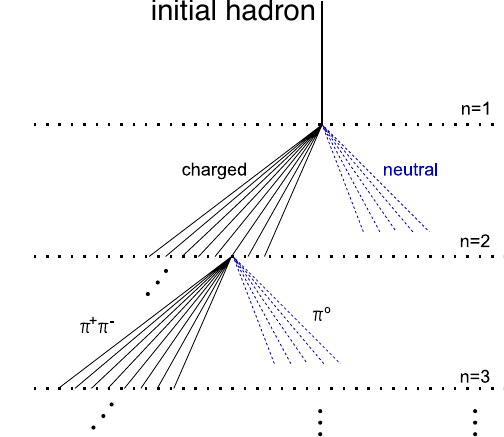}
	\caption{Simplified casecade model of air shower. The left panel shows the electromagnetic cascade and the right panel is the hadronic cascade from \cite{ulrich2011hadronic}. In the hadronic cascade, solid (dashed) lines represent charged (neutral) pions.}
	\label{fig.2}
\end{figure}

Extensive air showers (EAS) develop in complex processes as combination of electromagnetic cascades and hadronic cascades. It is necessary to perform detailed numerical simulations of air showers to infer the properties of the primary cosmic rays that initiate them. Thus, Heitler presented a simple model of electromagnetic(EM) cascade development \cite[]{heitler1984quantum}. The electromagnetic cascade involves two physical processes: the $e^\pm$ pairs production by photons and the bremsstrahlung by $e^\pm$. As depicted in the left panel of \autoref{fig.2}, high energy photons interact with atmospheric particles to produce $e^\pm$ pairs. The $e^\pm$ pairs continue to interact with nuclei during the propagation to produce bremsstrahlung, which generate photons. This cascade of photons and electrons along the shower axis is referred to as the electromagnetic cascade. Matthews generalised this concept of the Heitler model to hadronic interactions \cite[]{matthews2005heitler}. As sketched in the right panel of \autoref{fig.2}, the hadronic cascade in basic Heitler-Matthews model is approximated by a pure pion shower. Cosmic ray protons with an primary energy of $E_0$ collide with nuclei in the atmosphere, producing secondary high-energy hadrons. Neutral pions, with short lifetimes, decay into photons, leading to electromagnetic cascades. Charged pions continue to interact with particles in the atmosphere, producing hadronic interactions. Through a fixed amount of slant lengths, more $\pi^\pm$ and $\pi^0$ are produced until the energy of $\pi^{\pm}$ reach the critical energy $E_c^{\pi^\pm}\sim115\mathrm{GeV}$ \cite[]{thunman1996charm}. At this point, $\pi^{\pm}$ no longer undergo interactions, and the hadronic cascade stops. Then, all charged pions decay into muons at a 1:1 ratio. Expect pions, kaons can also decay into muons, with the critical energies $E_c^{K^\pm}\sim850\mathrm{GeV}$, $E_c^{K^0_L}\sim210\mathrm{GeV}$, and $E_c^{K^0_S}\sim30\mathrm{TeV}$ \cite[]{thunman1996charm}. The decay processes of $\pi$ and $K$ are shown in Eq.8.

\begin{equation}
	\begin{aligned}
		\pi^{\pm}&\rightarrow\mu^{\pm}+\nu_{\mu}/\bar{\nu}_{\mu}\\
		K^{+}&\rightarrow\pi^{+}\pi^{-},K^{+}\rightarrow\mu^{+}\nu_{\mu}\\
		K^{0}_{S}&\rightarrow\pi^{+}\pi^{-}\\
		K^{0}_{L}&\rightarrow\pi^{\pm}e^{\mp}\nu_e,K^{0}_{L}\rightarrow\pi^{\pm}\mu^{\mp}\nu_\mu.
	\end{aligned}
	\label{eq.8}
\end{equation}

Thus, the number of muons corresponds to the number of charged pions and kaons at the end of the hadronic cascade. In the basic Heitler-Matthews model, assuming the production number of pions up in each interaction is $n_{\mathrm{mult}}$, among which the fraction of charged pions is $r=2/3$ \cite[]{albrecht2022muon}. Thus, after undergoing $n_c$ critical cascade interactions, the total number of charged pions $N_{\mathrm{ch}}$, which is equal to the number of muons $N_{\mu}$, is given by $N_{\mu}=N_{\mathrm{ch}}=(rn_{\mathrm{mult}})^{n_{c}}$. When the energy of the charged pions equals the critical energy $E_c$, the cascade reaches the critical cascade generation $n_c$,  $N_{\mu}$ can be caclulated as:

\begin{equation}
	\begin{aligned}
		N_{\mu}=(rn_{\mathrm{mult}})^{n_{c}}=(\frac{E_{0}}{E_{c}})^{\beta}, 
	\end{aligned}
	\label{eq.9}	
\end{equation}
where $E_0$ is the primary energy of cosmic rays, and $\beta=\frac{\ln(rn_{\mathrm{mult}})}{\ln n_{\mathrm{mult}}}$. According to the Heitler-Matthews model, each generation of particles includes $\pi^0$ carrying one-third of the energy into the electromagnetic cascade. Taking into account the hadronic interactions and decay of kaons, as well as some other hadronic resonances, approximately $25\%$ of the energy in each generation of particles is transferred to the electromagnetic cascade \cite[]{thunman1996charm}. Finally, the electromagnetic cascade consumes about $90\%$ of the primary particle energy, while the remaining $10\%$ is carried by muons and neutrinos. 

The production of muon is a result of the secondary particles decay from hadronic cascades, so an increase of the energy proportion of the hadronic cascades will enhance the muon production. Therefore, it is necessary to focus on reducing the production of pion since $1/3$ of its energy will transfer to electromagnetic cascade through $\pi^{0} \rightarrow 2\gamma$ decay. Kaons (such as $K^+$, $K^-$, $K^0_S$, $K^0_L$) participate in hadronic interactions, as in Eq.\ref{eq.8}, after reaching the critical energy, their decay products include not only muons but also pions. The branching ratio of kaon decaying into $\pi^0$ is much lower than decaying into muon, as reported in \cite{patrignani2016review}. Thus, in this view, increasing the production of secondary particle kaons can increase the number of muons in the simulation after p-p collisions. Since kaons consist of a strange quark and a light quark, increasing the production of strange quarks is a key factor considered in this study.

\subsection{Production of pions and kaons in high-energy collisions}
\label{sec:3.2}
In the earlier discussion, we analyzed the possibility of gluon condensation (GC) phenomenon occurring in ultrahigh energy collisions. In fact, it has long been speculated that a new phase of matter, known as quark-gluon plasma (QGP), might exist in systems composed of deconfined quarks and gluons at high temperatures and densities \cite[]{rafelski2016melting}. In high-energy collidings of two nuclei, the energy density and temperature in the central rapidity region may be high enough to potentially form quark-gluon plasma . In such condition, the ground state of matter is the quark-gluon plasma phase (QGP) rather than the hadronic phase. In ordinary nuclear matter, the content of strange quarks $s$($\bar{s}$ ) are much less than that of $u$$\bar{u}$ and $d$$\bar{d}$. In nucleon-nucleon collisions, $u(\bar{u}\,),\, d(\bar{d}\,)$ and $s$($\bar{s}$ ) are produced, followed by the combination of strange quarks( anti-quarks) with neighboring light anti-quarks (quarks) to form strange particles. The appearance of quark-gluon plasma leads to an increase number of strange quarks and anti-quarks. The occurrence of the GC requires a higher collision energy and we find that it  leads to  more strange quarks (anti-quarks) production than in the condition of  QGP. Next, we will first discuss the production of strange particles in the QGP.

In the high energy collisions, $s\bar{s}$ pairs can be produced through two processes: $q+\bar{q}\rightarrow s+\bar{s}$ and $g+g\rightarrow s+\bar{s}$, where $q$ is light quarks $u$ or $d$. The scattering cross-sections for these two reactions are given by Eq.10 \cite[]{gluck1978duality,combridge1979associated,matsui1986strangeness} and Eq.11 \cite[]{gluck1978duality}, respectively.

\begin{equation}
	\sigma_{q\bar{q}}(s)=\frac{8\pi\alpha_{s}^{2}}{27s}(1+\frac{\eta}{2})\sqrt{1-\eta},
	\label{eq.10}
\end{equation}

\begin{equation}			
	\sigma_{gg}(s)=\frac{\pi\alpha_{s}^{2}}{3s}\left[(1+\eta+\frac{1}{16}\eta^{2})\ln\left(\frac{1+\sqrt{1-\eta}}{1-\sqrt{1-\eta}}\right)-\left(\frac{7}{4}+\frac{31}{16}\eta\right)\sqrt{1-\eta}\right],
	\label{eq.11}
\end{equation}
in the two equations, $s=\left(q+\bar{q}\right)^2$ or $s=\left(g+g\right)^2$ is the square of center-of-mass energy of the corresponding process. $\eta={4m_s^2}/s$, where $m_s$ is the mass of the strange quark and its antiquark. The reaction will not appear if the energy $\sqrt{s}$ is less than the threshold energy $2m_s$. In the QGP state, the momentum distribution functions of quarks and gluons follow the Fermi-Dirac and Bose-Einstein distributions, respectively.

\begin{equation}
	\begin{aligned}
		&f_{q}(k_{1})=\frac{1}{\mathrm{e}^{k_1-\mu/T}+1}\\
		&f_{\overline{q}}(k_{2})=\frac{1}{e^{k_2+\mu/T}+1}\\
		&f_{g}(k)=\frac{1}{\mathrm{e}^{k/T}-1},
		\label{eq.12}
	\end{aligned}
\end{equation}
the four-momentum of the incident particles is denoted by $k_i$, and $\mu$ is the quark chemical potential. The total yield of $s\bar{s}$ pairs per unit space-time volume is given by:

\begin{equation}
	\begin{aligned}
		A=\frac{\mathrm{d}N}{\mathrm{d}t\mathrm{d}^3x}&=\frac{1}{2}\int_{4m_s^2}^\infty s\mathrm{d}s\sigma(s-(k_1+k_2)^2)\\
		&\times\left\{N_{c}^{2}N_{s}^{2}N_{f}\int\frac{\mathrm{d}^{3}k_{1}\mathrm{d}^{3}k_{2}}{(2\pi)^{6}|k_1||k_2|}f_{q}(k_{1})f_{\bar{q}}(k_{2})\sigma_{q\bar{q}}(s)+N_{g}^{2}N_{\epsilon}^{2}\int\frac{\mathrm{d}^{3}k_{1}\mathrm{d}^{3}k_{2}}{(2\pi)^{6}|k_1||k_2|}f_{g}(k_{1})f_{g}(k_{2})\sigma_{gg}(s)\right\}.
	\end{aligned}
	\label{eq.13}
\end{equation}
where $N_c=3$ is the color number of quarks, $N_g=8$ is the color number of gluons, $N_\epsilon=2$ is the number of polarizations, $N_f=2$ is the flavor degeneracy in QGP. For the process $q+\bar{q}\rightarrow s+\bar{s}$, we have $N_{q\bar{q}}=N_{c}^{2}N_{s}^{2}N_{f}=72$; for the process $g+g\rightarrow s+\bar{s}$, we have $N_{gg}=N_{g}^{2}N_{\epsilon}^{2}=256$. It can be seen that the total production rate of strangeness density is mainly dominated by gluon collisions. Many experimental collaborations, such as CERN's NA34 \cite[]{van1991na34,mazzoni1992measurements} and BNL's E802 \cite[]{nagamiya1992experimental,abbott1991comparison,E802:1992wow}, have conducted research on strangeness in high-energy heavy-ion collisions. In condition of the incident particle energy at the scale of GeV, $n_{K^+}/n_{\pi^+}$ can reach approximately 0.24 \cite[]{wong1994introduction}. In condition of thermal and chemical equilibrium at $T=200$ MeV, the production of pions and kaons in quark-gluon plasma state will be $n_{K^+}/n_{\pi^+} \approx 0.38$. Although the strangeness increases, it's still not rich enough to solve the problem of muon excess \cite[]{anchordoqui2017strange}.

Substituting the gluon distribution function Eq.\ref{eq.12} in QGP state into Eq.\ref{eq.13}, one get the yield in the quark-gluon plasma state \cite[]{rafelski1982strangeness}, 

\begin{equation}
	A_{\mathrm{QGP}}=\frac{51}{3\pi^2}\alpha_s^2m_sT^3e^{-2m_s/T}(1+\frac{51T}{14m_s}+\cdots).
	\label{eq.14}
\end{equation}

In GC conditions, the distribution of gluons tends to be a delta function like, 

\begin{equation}
	f_g\left(k\right)=C_m\delta(k-k_c)
	\label{eq.15}
\end{equation}
where $k_c=\frac{3}{2}T$ and $C_m\textless1.0$, we take $C_m=0.4$ (a lower limit) based on the QCD calculations and global fitting of gluon distributions \cite[]{gao2018structure}. 
Substituting Eq.\ref{eq.15} into Eq.\ref{eq.13}, we obtain,

\begin{equation}
	A_{\mathrm{GC}}=C_m^2\frac{128\alpha_s^2k_c^4}{3\pi^3}\left(\mathrm{log} \frac{4k_c^2}{m_s^2}-1\right).
	\label{eq.16}
\end{equation}

Figure.3 shows the comparison of $A_{\mathrm{GC}}$ and $A_{\mathrm{QGP}}$ as functions of $T$. The parameters used here are $\alpha_s=2.2$ and $m_s=280$ MeV \cite[]{rafelski1982strangeness}.

\begin{figure}%[H] 
	\centering
	\includegraphics[width=0.6\textwidth]{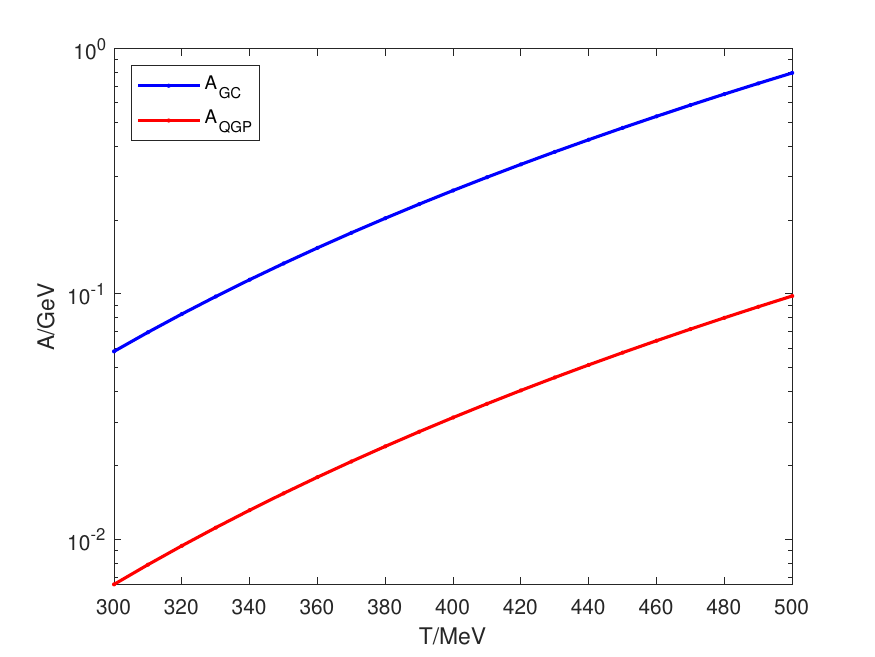}
	\caption{Comparison of the production rate of strange quark pairs per unit spacetime volume in the Gluon Condensate (GC) state and the Quark-Gluon Plasma (QGP) state.}	
	\label{fig.3}
\end{figure}

It's evident that the value of $A_{\mathrm{GC}}$ is significantly greater than $A_{\mathrm{QGP}}$ in \autoref{fig.3} with $A_{\mathrm{GC}}/A_{\mathrm{OGP}}\approx 8.9$. The production rates of strange quarks in both conditions are positively correlated with temperature. The trends of the two curves appear similar in the plot, and the value of $A_{\mathrm{GC}}$ shows much larger than $A_{\mathrm{QGP}}$ within the range of temperature $T$. This indicates that the production rate of strange quarks in GC condition is much richer than that in QGP condition. So, a significant enhancement in the number of kaons in the hadronization products can be expected in the gluon condensation state.

\subsection{Productions of Pions and Kaons in the GC Model}
\label{sec:3.3}
In this section, we will introduce the production of pions and kaons in the gluon condensation model. In this model,  the distribution of gluons at the critical momentum $k_c$ is shown in Eq.\ref{eq.15}. When the energy of colliding particles in the center of momentum system reaches the energy threshold, gluon condensation will occur. In our model, the first stage of air shower cascade, initiated by ultrahigh-energy cosmic rays (taking $E_0=10^{19}$ eV as an example), reach the energy threshold for the occurrence of gluon condensation. For simplicity, we assume that only pions and kaons are the new productions.
In condition of GC state, abundant strange quarks are generated from the interactions between the huge number of gluons ($g+g\rightarrow s+\bar{s}$). In the subsequent hadronization processes, the number of kaons  increases significantly. 
Kaons include $K^+$, $K^-$, $K^0_L$, and $K^0_S$. At the EeV energy scale, the isospin symmetry breaking in kaons can be ignored \cite[]{NA61SHINE:2023azp}, so the four types of kaons in this paper are assumed to be produced equally.

In hadron-hadron collisions,  the leading particles (similar to the  incident and target particles) in the fragmentation region are the fast-moving particles in the forward or backward directions, which carry a significant portion of the energy at the center-of-mass system. The degree of inelasticity of the collision is usually described by the forward light-cone variable $x$. For the reaction $b+a\rightarrow c+X$, $x=\frac{c_0+c_z}{b_0+b_z}$, where $c_0$ and $b_0$ are the energies of particles $c$ and $b$, and $c_z$ and $b_z$ are the longitudinal momenta. The measured cross-section distribution $\mathrm{d}\sigma/\mathrm{d}x$ as a function of the light-cone variable $x$ reflects the degree of energy loss in inelastic collisions. Except in the diffraction dissociation region where the light-cone $x\sim 1$, the distribution is nearly flat. If extrapolated to the small $x$ region, excluding diffraction dissociation, the $\mathrm{d}\sigma/\mathrm{d}x$ can be given as:

\begin{equation}
	\mathrm{d}\sigma/\mathrm{d}x\approx\sigma_{\mathrm{in}}\theta(1-x)\theta(x-x_{\mathrm{L}})/(1-x_{\mathrm{L}})
	\label{eq.17}
\end{equation}
where $\sigma_{\mathrm{in}}$ is the total inelastic cross-section, and $x_\mathrm{L}$ is the lower limit of $x$. The distribution in Eq.\ref{eq.17} indicates that after the inelastic collisions, the mean $x$ is about $1/2$, implying that about half of the initial baryons' light-cone momentum is carried away.

\begin{figure}[htbp]
	\centering 
	\includegraphics[width=0.5\textwidth]{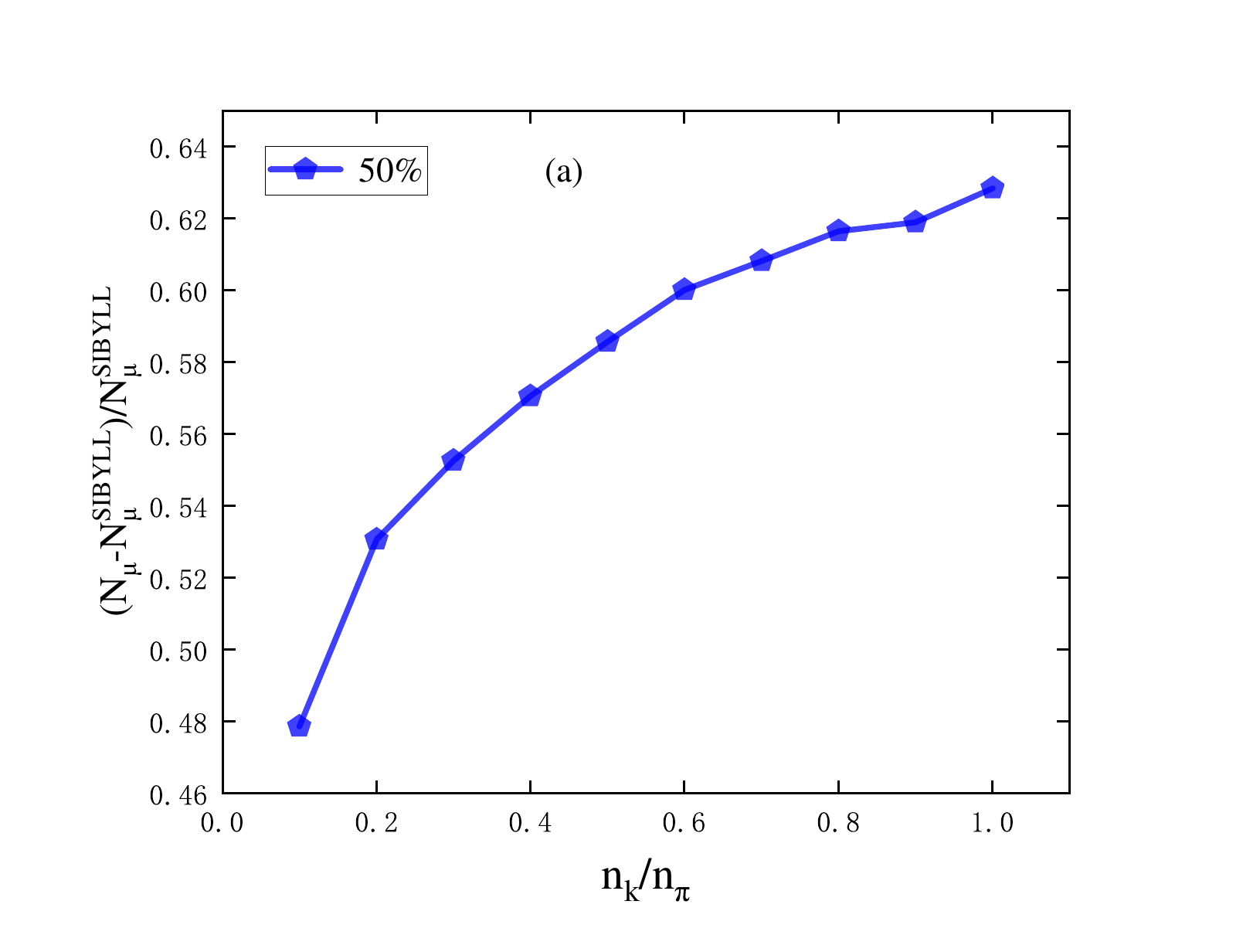}\hfill
	\includegraphics[width=0.5\textwidth]{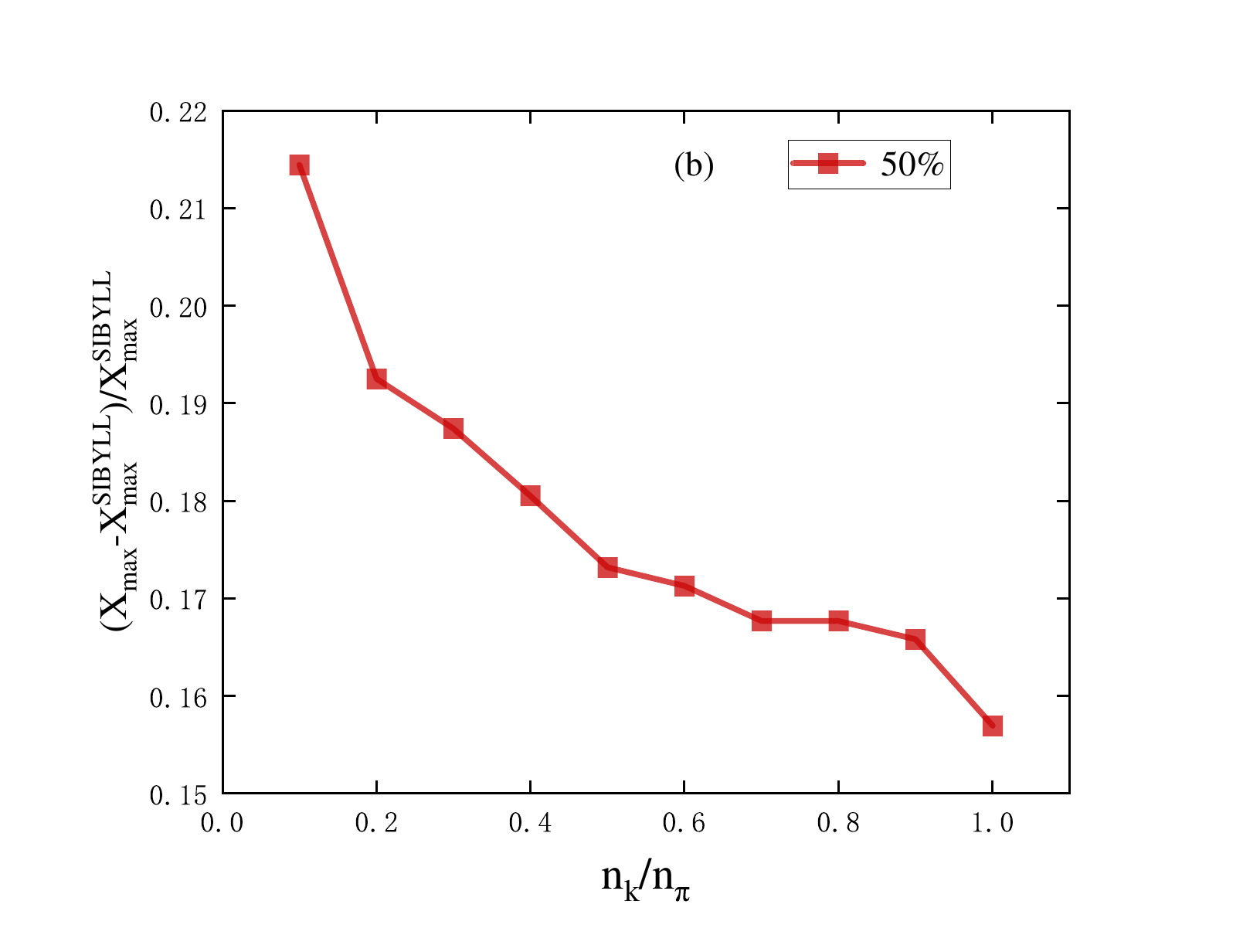}
	
	\includegraphics[width=0.5\textwidth]{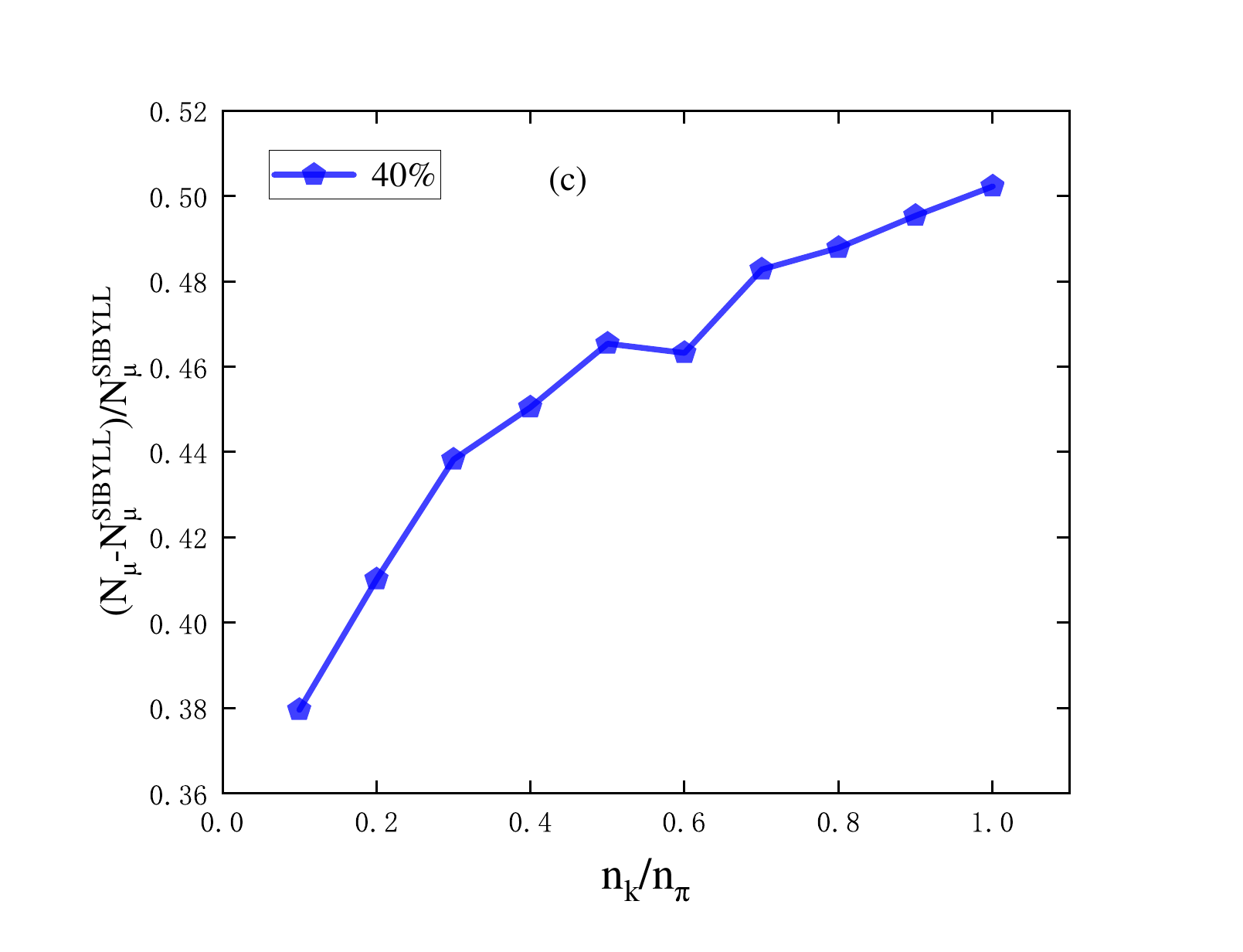}\hfill
	\includegraphics[width=0.5\textwidth]{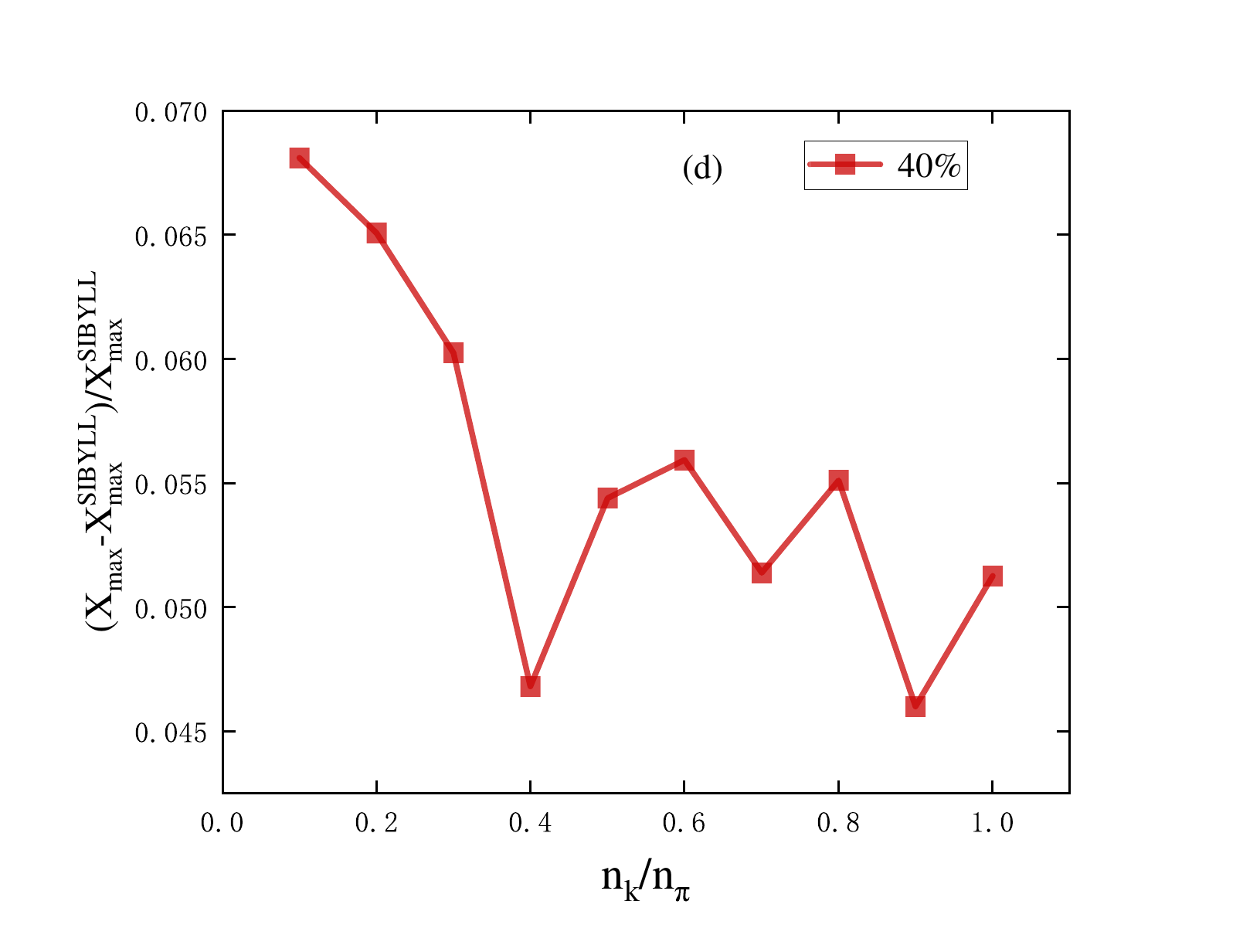}
	
	\includegraphics[width=0.5\textwidth]{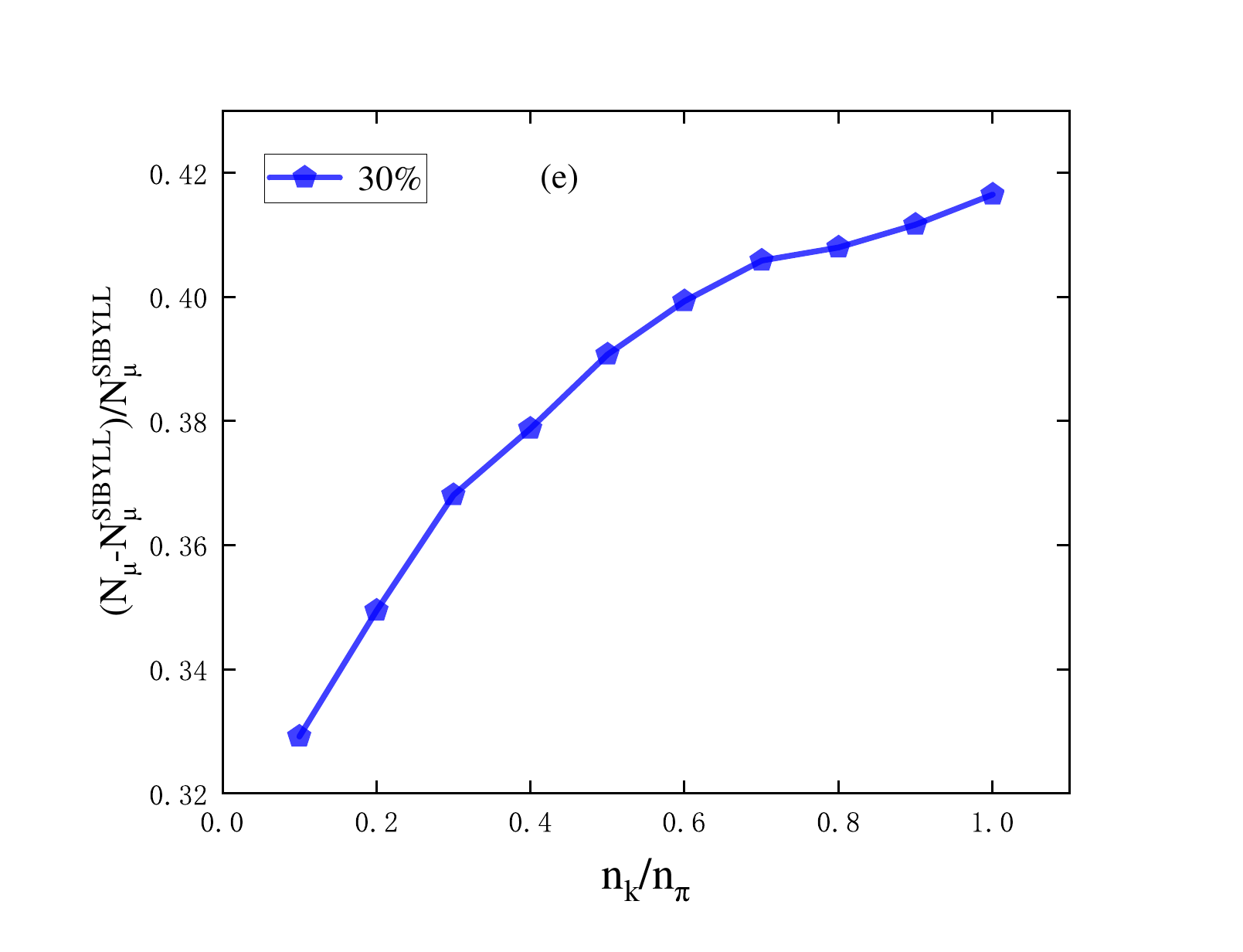}\hfill
	\includegraphics[width=0.5\textwidth]{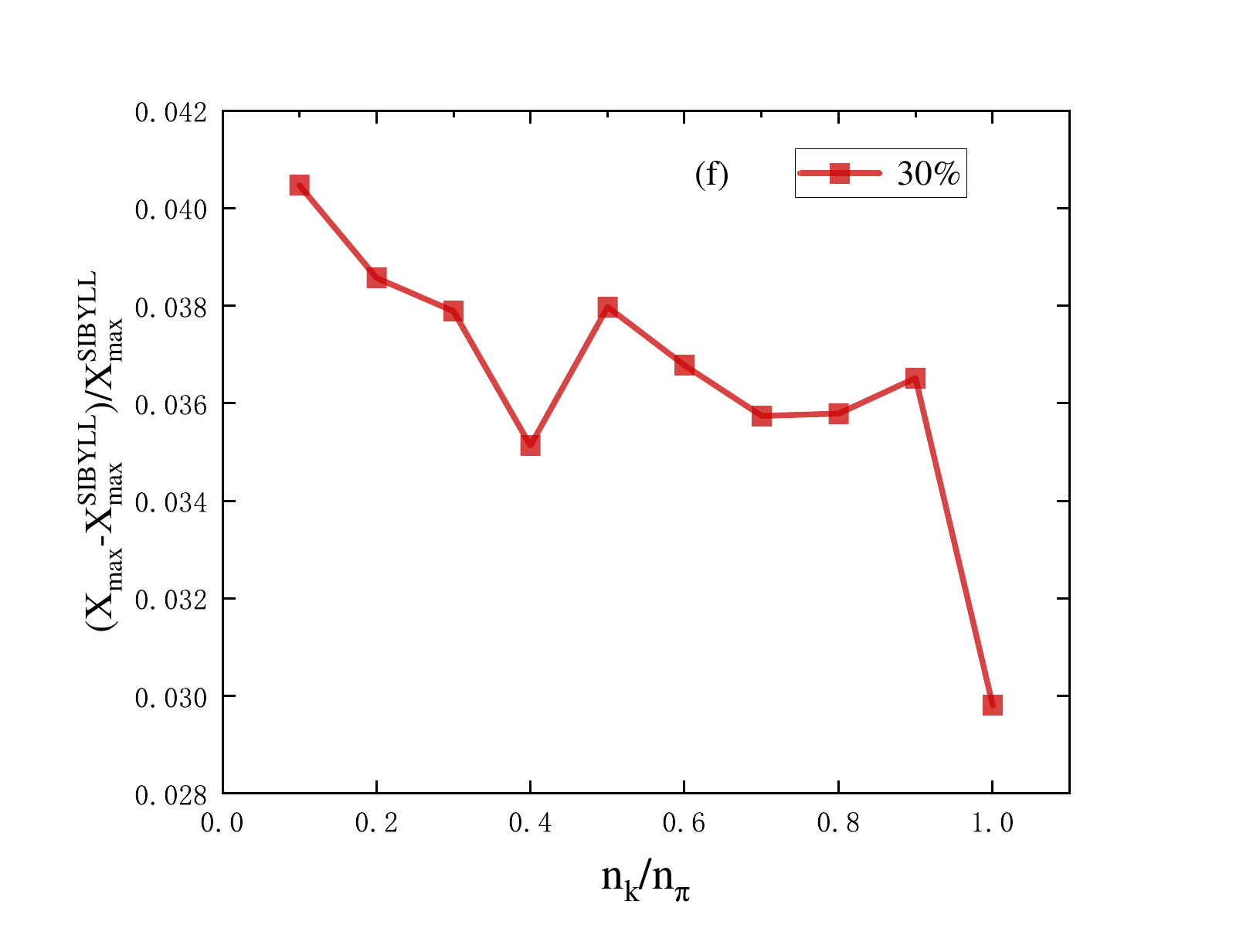}
	\caption{In the left panel, $(N_\mu-N_\mu^{\mathrm{SIBYLL}})/N_\mu^{\mathrm{SIBYLL}}$  changing with $n_K/n_\pi$ are shown, where $n_K/n_\pi$ corresponding to the parameter $c$ in Eq.\ref{eq.6}. $N_\mu$ and $N_\mu^{\mathrm{SIBYLL}}$ are the results of our model and  the hadronic model SIBYLL \cite[]{engel2017hadronic} seperatly.  In the right panel, $(X_\mathrm{max}-X_\mathrm{max}^{\mathrm{SIBYLL}})/X_\mathrm{max}^{\mathrm{SIBYLL}}$ are shown. $X_\mathrm{max}$ and $X_\mathrm{max}^{\mathrm{SIBYLL}}$ are the depth of shower maximum of our model and  the model SIBYLL. The parameter $50\%, 40\%,30\%$ in the figure corresponding to the energy proportion taken by pions and kaons.}
	\label{fig.4}
\end{figure}

In this study, the theoretical framework outlined in Eq.\ref{eq.4} to Eq.\ref{eq.8} serves as the basis to simulate the secondary particle products resulting from showers in proton-proton collisions in various conditions. We consider three conditions seperatly, that is , the leading protons, carrying away proportions of the primary energy with $50\%$, $60\%$ and $70\%$, which means that the total energy carried by pions and kaons are $50\%$, $40\%$, and $30\%$ correspondingly.
These simulations are compared with the traditional result derived from the SIBYLL model \cite[]{fletcher1994s,engel2017hadronic}. In \autoref{fig.4}, the simulated results in the GC condition for pions and kaons are shown. They illustrate the relative increases of $N_\mu$ and $X_{\mathrm{max}}$ compared to the traditional model, along with the $n_K/n_\pi$. The relative increases are quantified as $(N_\mu-N_\mu^{\mathrm{SIBYLL}})/N_\mu^{\mathrm{SIBYLL}}$ and $(X_\mathrm{max}-X_\mathrm{max}^{\mathrm{SIBYLL}})/X_\mathrm{max}^{\mathrm{SIBYLL}}$. 
There are some key characteristics of the results: 

(1) The simulation results primarily based on the GC model exhibit a significant enhancement in EAS of $N_\mu$, surpassing the traditional model by more than $30\%$, as shown in the left part of Figure 4. For a single figure, for example Figure 4(a), as $n_K/n_\pi$  increases, $(N_\mu-N_\mu^{\mathrm{SIBYLL}})/N_\mu^{\mathrm{SIBYLL}}$ also gradually increases, that means we can increase the muon production by increasing the  $n_K/n_\pi$. On the other hand, comparing  Figure 4(a), (c) and (e), we also find that the excess proportion increases with the rise in the energy fraction carried by pions and kaons. Particularly in Figure 4(a), the excess proportion can reach up to $60\%$ within the specified range of $n_K/n_\pi$.

(2)  The energy fraction carried by pions and kaons  has a significant impact on the depth of shower maximum $X_{\mathrm{max}}$. As shown in the right part of Figure 4, as the energy fraction decreases, the $(X_\mathrm{max}-X_\mathrm{max}^{\mathrm{SIBYLL}})/X_\mathrm{max}^{\mathrm{SIBYLL}}$ decreases. It is noteworthy that the ordinate value in Figure 4(b) is about $0.16\sim 0.21$, indicating a notable deviation between the simulated  $X_{\mathrm{max}}$ and the experimental results under the condition where the secondary particles pions and kaons carrying about $50\%$ of the primary energy. In Figure 4(d) and (f), $(X_\mathrm{max}-X_\mathrm{max}^{\mathrm{SIBYLL}})/X_\mathrm{max}^{\mathrm{SIBYLL}}$ is significantly decreased.

(3) Since the SIBYLL model does not indicate obvious deviation between simulated $X_\mathrm{max}$ and the experimental data, we can conclude that the  pions and kaons carring $50\%$ of the total energy (Figure 4(b)) is not a suitable choice. In Figure 4(d) and (f), the relative deviation is within the range of $0.07$ and $0.04$. 
Considering the corresponding experimental error of $X_\mathrm{max}$, it is acceptable to take the values $40\%$ or $30\%$ of the primary energy carried by mesons as in Figure 4(d) and (f). 
A better balance can be achieved between solving the problem of muon excess and maintaining the consistency with the depth of shower maximum.  

\begin{figure}[htbp]
	\centering 
	\includegraphics[width=0.5\textwidth]{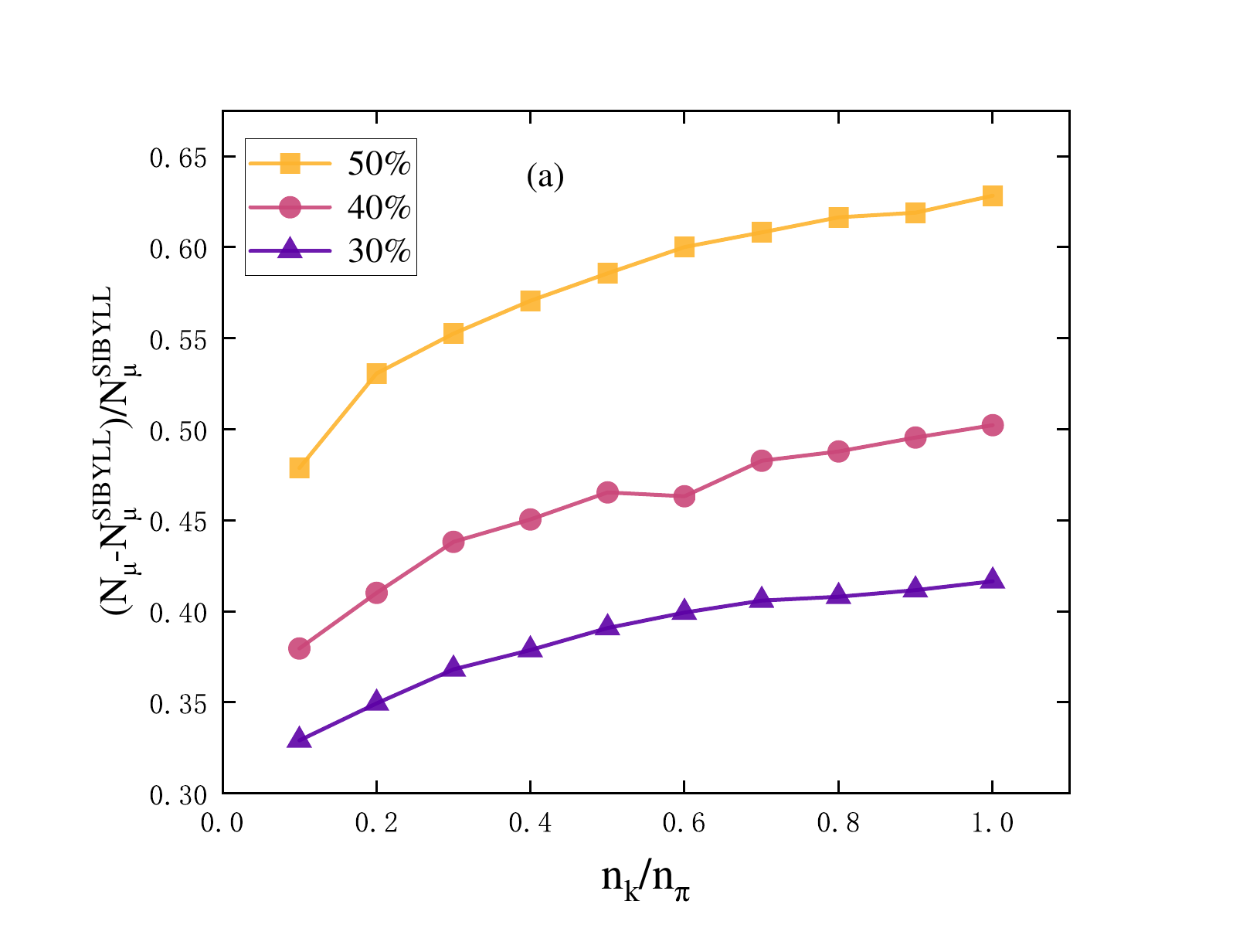}\hfill
	\includegraphics[width=0.5\textwidth]{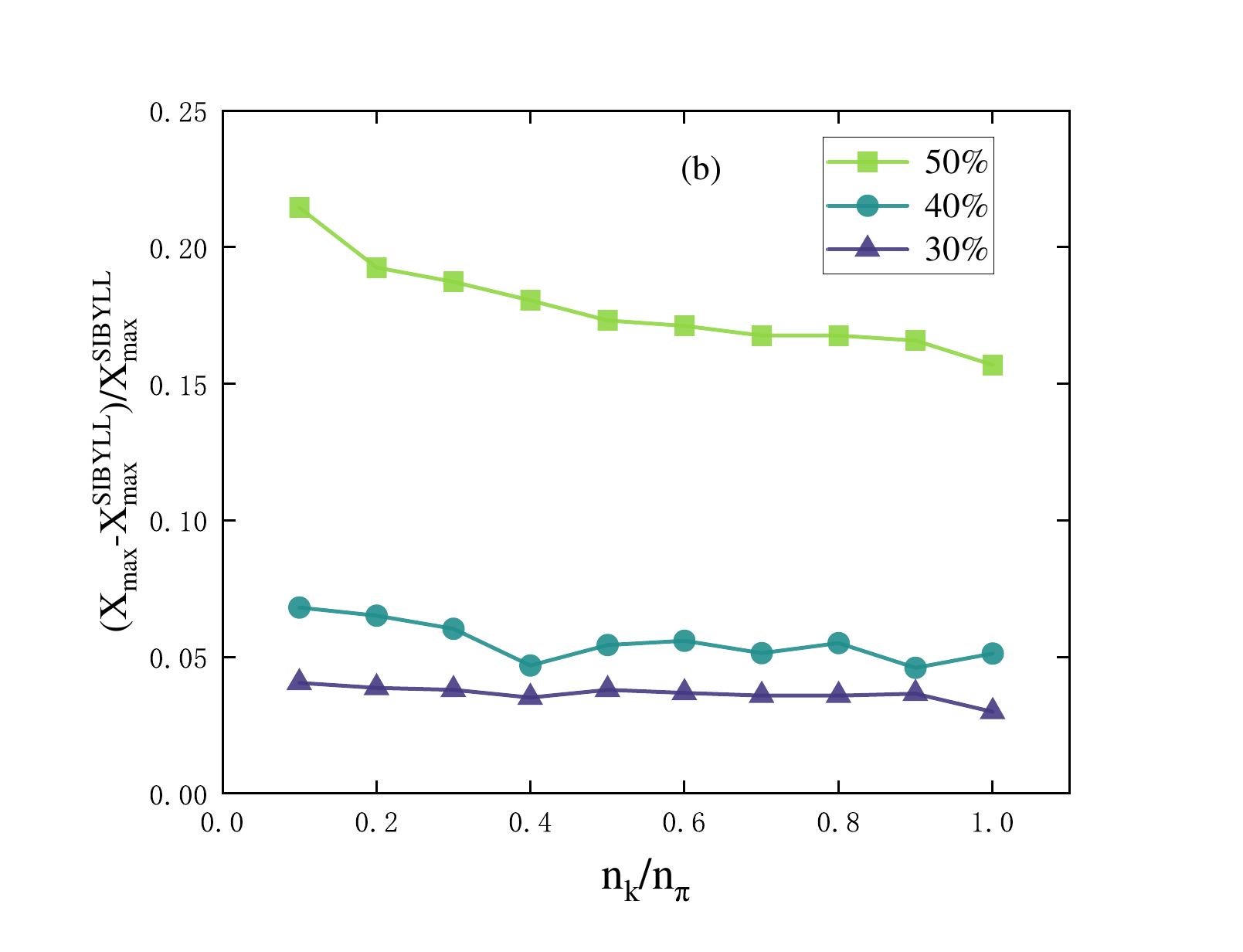}
	
	\caption{(a) The relative increase in $N_\mu$ with different primary energy distributions. (b) The relative increase in $X_\mathrm{max}$.}
	\label{fig.5}
\end{figure}

\autoref{fig.5} merges the results of different energy ratios from \autoref{fig.4} for intuitive comparison.
Drawing insights from Figure 5(a), it is evident that the occurrence of the GC effect during the collision process leads to a significant increase in the muon yield. Moreover, as the value $n_K/n_\pi$ increases, the predicted muon excess becomes more pronounced. It is important to emphasize that the increase of $N_\mu$ should be consistent with the experimental results of the depth of shower maximum $X_\mathrm{max}$. Thus, by examing the results in Figure 5(b), we deduce that the energy carried by the leading protons should exceed $60\%$ of the primary energy to provide a more comprehensive explanation for the observed muon excess phenomenon.

\section{Discussion and conclusion}
\label{sec:4}

In high-energy collision experiments, the existing research suggest that QGP may occur, but considering only the conventional QGP effect cannot solve the muon excess problem in air shower. We consider that GC may occur in extremely high energy collisions. In the GC state, the production of strange quarks will be greatly enhanced, so that the $n_K/n_\pi$ can far exceed the $n_K/n_\pi \approx 0.24$ in ordinary situations and $n_K/n_\pi \approx 0.38$ in QGP, reaching $n_K/n_\pi>1.0$.  In figure.4 and  figure.5, we simulated $(N_\mu-N_\mu^{\mathrm{SIBYLL}})/N_\mu^{\mathrm{SIBYLL}}$ and $(X_\mathrm{max}-X_\mathrm{max}^{\mathrm{SIBYLL}})/X_\mathrm{max}^{\mathrm{SIBYLL}}$ in the $0.1 <n_K/n_\pi< 1.0$ region under GC conditions and obtained reasonable results. 

According to \cite{anchordoqui2017strange}, when the energy of incident particle is about $10^{19}$ eV, the measured number of muons exceeds the theoretical prediction by approximately $40\%$. 
In this study, considering only the gluon condensation effect in the proton-proton collisions, the results in figure.\ref{fig.4} indicate a significant enhancement in $N_\mu$ in air showers. The muon number increases with the value of $n_K/n_\pi$ increasing. For $n_K/n_\pi=0.6$, the production of muon in the GC model reaches the expected enhancement of $40\%$ when the mesons taken about $30\%\sim40\%$ of the incident energy. When $n_K/n_\pi>0.6$, the increase becomes relatively slow. Considering that the results should not conflict with the measured $X_\mathrm{max}$ simultaneously, the number of muons in our model can be increased by up to $50\%$ compared to the traditional models in condition of $n_K/n_\pi<1.0$. As pointed out by \cite{albrecht2022muon}, the muon excess phenomenon becomes more pronounced with the increasing primary energy. In GC model, considering that $n_K/n_\pi$ will increase as the the energy of incident cosmic rays increase, it will be naturally obtained.

Although we cannot determine how much energy the leading particle takes away, roughly speaking, as the energy of the incident particle increases, the proportion of the energy carried by the leading particle will increase. Therefore, in the case of GC, it is possible for the energy of the leading particle to exceed $60\%$. In this way, while explaining the muon excess problem, the impact on the depth of the maximum shower $X_\mathrm{max}$ is minimal, and there will be no contradiction with this experimental measurements.  Our work provides a possible explanation to the muon excess problem.

\section*{Acknowledgements}
This work is supported by the National Natural Science Foundation of China (No.11851303).

\bibliographystyle{aasjournal}

\begin{thebibliography}{}
\expandafter\ifx\csname natexlab\endcsname\relax\def\natexlab#1{#1}\fi
\providecommand{\url}[1]{\href{#1}{#1}}
\providecommand{\dodoi}[1]{doi:~\href{http://doi.org/#1}{\nolinkurl{#1}}}
\providecommand{\doeprint}[1]{\href{http://ascl.net/#1}{\nolinkurl{http://ascl.net/#1}}}
\providecommand{\doarXiv}[1]{\href{https://arxiv.org/abs/#1}{\nolinkurl{https://arxiv.org/abs/#1}}}

\bibitem[{Aab {et~al.}(2015)}]{PierreAuger:2014ucz}
Aab, A., {et~al.} 2015, \prd, 91, 032003, \dodoi{10.1103/PhysRevD.91.032003}

\bibitem[{Aab {et~al.}(2016{\natexlab{a}})}]{aab2016testing}
---. 2016{\natexlab{a}}, \prl, 117, 192001,
  \dodoi{10.1103/PhysRevLett.117.192001}

\bibitem[{Aab {et~al.}(2016{\natexlab{b}})}]{aab2016pierre}
---. 2016{\natexlab{b}}, arXiv e-prints, arXiv:1604.03637,
  \dodoi{10.48550/arXiv.1604.03637}

\bibitem[{Aab {et~al.}(2017)Aab, Abreu, Aglietta, Al~Samarai, Albuquerque,
  Allekotte, Almela, Castillo, Alvarez-Mu{\~n}iz, Anastasi,
  {et~al.}}]{aab2017combined}
Aab, A., Abreu, P., Aglietta, M., {et~al.} 2017, \jcap, 2017, 038,
  \dodoi{10.1088/1475-7516/2017/04/038}

\bibitem[{Aab {et~al.}(2021)}]{PierreAuger:2021qsd}
Aab, A., {et~al.} 2021, \prl, 126, 152002,
  \dodoi{10.1103/PhysRevLett.126.152002}

\bibitem[{Abbott {et~al.}(1991)Abbott, Akiba, Beavis, Bloomer, Bond, Chasman,
  Chen, Chu, Cole, Costales, {et~al.}}]{abbott1991comparison}
Abbott, T., Akiba, Y., Beavis, D., {et~al.} 1991, \prl, 66, 1567,
  \dodoi{10.1103/PhysRevLett.66.1567}

\bibitem[{Adhikary {et~al.}(2023{\natexlab{a}})}]{NA61SHINE:2022tiz}
Adhikary, H., {et~al.} 2023{\natexlab{a}}, \prd, 107, 062004,
  \dodoi{10.1103/PhysRevD.107.062004}

\bibitem[{Adhikary {et~al.}(2023{\natexlab{b}})}]{NA61SHINE:2023azp}
---. 2023{\natexlab{b}}, arXiv e-prints, arXiv:2312.06572,
  \dodoi{10.48550/arXiv.2312.06572}

\bibitem[{Adriani {et~al.}(2016)Adriani, Berti, Bonechi, Bongi, D'Alessandro,
  Del~Prete, Haguenauer, Itow, Iwata, Kasahara, Kawade, Makino, Masuda,
  Matsubayashi, Menjo, Mitsuka, Muraki, Papini, Perrot, Ricciarini, Sako,
  Sakurai, Suzuki, Tamura, Tiberio, Torii, Tricomi, Turner, Ueno, \&
  Zhou}]{PhysRevD.94.032007}
Adriani, O., Berti, E., Bonechi, L., {et~al.} 2016, \prd, 94, 032007,
  \dodoi{10.1103/PhysRevD.94.032007}

\bibitem[{Ahn {et~al.}(2009)Ahn, Engel, Gaisser, Lipari, \&
  Stanev}]{ahn2009cosmic}
Ahn, E.-J., Engel, R., Gaisser, T.~K., Lipari, P., \& Stanev, T. 2009, \prd,
  80, 094003, \dodoi{10.1103/PhysRevD.80.094003}

\bibitem[{Albrecht {et~al.}(2022)Albrecht, Cazon, Dembinski, Fedynitch,
  Kampert, Pierog, Rhode, Soldin, Spaan, Ulrich, {et~al.}}]{albrecht2022muon}
Albrecht, J., Cazon, L., Dembinski, H., {et~al.} 2022, Astrophys. Space Sci.,
  367, 27, \dodoi{10.1007/s10509-022-04054-5}

\bibitem[{Allen \& Farrar(2013)}]{allen2013testing}
Allen, J., \& Farrar, G. 2013, arXiv e-prints, arXiv:1307.7131,
  \dodoi{10.48550/arXiv.1307.7131}

\bibitem[{Altarelli \& Parisi(1977)}]{altarelli1977asymptotic}
Altarelli, G., \& Parisi, G. 1977, Nuclear Physics B, 126, 298,
  \dodoi{10.1016/0550-3213(77)90384-4}

\bibitem[{Anchordoqui {et~al.}(2022)Anchordoqui, Canal, Kling, Sciutto, \&
  Soriano}]{anchordoqui2022explanation}
Anchordoqui, L.~A., Canal, C.~G., Kling, F., Sciutto, S.~J., \& Soriano, J.~F.
  2022, JHEAp, 34, 19, \dodoi{10.1016/j.jheap.2022.03.004}

\bibitem[{Anchordoqui {et~al.}(2017)Anchordoqui, Goldberg, \&
  Weiler}]{anchordoqui2017strange}
Anchordoqui, L.~A., Goldberg, H., \& Weiler, T.~J. 2017, \prd, 95, 063005,
  \dodoi{10.1103/PhysRevD.95.063005}

\bibitem[{Balitsky \& Lipatov(1978)}]{balitsky1978pomeranchuk}
Balitsky, Y.~Y., \& Lipatov, L. 1978, Yad. Fiz., 28, 1597

\bibitem[{Baur {et~al.}(2023)Baur, Dembinski, Perlin, Pierog, Ulrich, \&
  Werner}]{baur2023core}
Baur, S., Dembinski, H., Perlin, M., {et~al.} 2023, \prd, 107, 094031,
  \dodoi{10.1103/PhysRevD.107.094031}

\bibitem[{Chang {et~al.}(2014)Chang, Mezrag, Moutarde, Roberts,
  Rodr{\'\i}guez-Quintero, \& Tandy}]{chang2014basic}
Chang, L., Mezrag, C., Moutarde, H., {et~al.} 2014, Physics Letters B, 737, 23,
  \dodoi{10.1016/j.physletb.2014.08.009}

\bibitem[{Combridge(1979)}]{combridge1979associated}
Combridge, B. 1979, Nuclear Physics B, 151, 429,
  \dodoi{10.1016/0550-3213(79)90449-8}

\bibitem[{Dokshitzer(1977)}]{dokshitzer1977calculation}
Dokshitzer, Y.~L. 1977, Journal of Experimental and Theoretical Physics, 46,
  641

\bibitem[{Enberg \& Peschanski(2006)}]{Enberg:2005zj}
Enberg, R., \& Peschanski, R.~B. 2006, Nucl. Phys. A, 767, 189,
  \dodoi{10.1016/j.nuclphysa.2005.12.012}

\bibitem[{Engel {et~al.}(2017)Engel, Riehn, Fedynitch, Gaisser, \&
  Stanev}]{engel2017hadronic}
Engel, R., Riehn, F., Fedynitch, A., Gaisser, T.~K., \& Stanev, T. 2017, EPJ
  Web Conf., 145, 08001, \dodoi{10.1051/epjconf/201614508001}

\bibitem[{Fadin {et~al.}(1975)Fadin, Kuraev, \& Lipatov}]{fadin1975pomeranchuk}
Fadin, V.~S., Kuraev, E., \& Lipatov, L. 1975, Physics Letters B, 60, 50,
  \dodoi{10.1016/0370-2693(75)90524-9}

\bibitem[{Fletcher {et~al.}(1994)Fletcher, Gaisser, Lipari, \&
  Stanev}]{fletcher1994s}
Fletcher, R., Gaisser, T., Lipari, P., \& Stanev, T. 1994, \prd, 50, 5710,
  \dodoi{10.1103/PhysRevD.50.5710}

\bibitem[{Gaisser {et~al.}(2016)Gaisser, Engel, \& Resconi}]{gaisser2016cosmic}
Gaisser, T.~K., Engel, R., \& Resconi, E. 2016, {Cosmic rays and particle
  physics}, 1st edn. (Cambridge University Press)

\bibitem[{Gao {et~al.}(2018)Gao, Harland-Lang, \& Rojo}]{gao2018structure}
Gao, J., Harland-Lang, L., \& Rojo, J. 2018, Physics Reports, 742, 1,
  \dodoi{10.1016/j.physrep.2018.03.002}

\bibitem[{Gl{\"u}ck \& Reya(1978)}]{gluck1978duality}
Gl{\"u}ck, M., \& Reya, E. 1978, Physics Letters B, 79, 453,
  \dodoi{10.1016/0370-2693(78)90405-7}

\bibitem[{Gribov \& Lipatov(1972)}]{gribov1972deep}
Gribov, V.~N., \& Lipatov, L.~N. 1972, {DEEP INELASTIC ep-SCATTERING IN A
  PERTURBATION THEORY.}, Tech. rep., Inst. of Nuclear Physics, Leningrad

\bibitem[{Heitler(1984)}]{heitler1984quantum}
Heitler, W. 1984, The quantum theory of radiation, 3rd edn. (Courier
  Corporation)

\bibitem[{Kuraev {et~al.}(1976)Kuraev, Lipatov, \&
  Fadin}]{kuraev1976multireggeon}
Kuraev, E.~A., Lipatov, L., \& Fadin, V.~S. 1976, Zhurnal Ehksperimental'noj i
  Teoreticheskoj Fiziki, 71, 840

\bibitem[{Mathieu \& Vento(2010)}]{mathieu2010pos}
Mathieu, V., \& Vento, V. 2010, proceedings of science

\bibitem[{Matsui {et~al.}(1986)Matsui, Svetitsky, \&
  McLerran}]{matsui1986strangeness}
Matsui, T., Svetitsky, B., \& McLerran, L.~D. 1986, \prd, 34, 783,
  \dodoi{10.1103/PhysRevD.34.783}

\bibitem[{Matthews(2005)}]{matthews2005heitler}
Matthews, J. 2005, \ap, 22, 387, \dodoi{10.1016/j.astropartphys.2004.09.003}

\bibitem[{Mazzoni {et~al.}(1992)Mazzoni, Collaboration,
  {et~al.}}]{mazzoni1992measurements}
Mazzoni, M., Collaboration, H., {et~al.} 1992, Nuclear Physics A, 544, 623,
  \dodoi{10.1016/0375-9474(92)90630-3}

\bibitem[{Nagamiya(1992)}]{nagamiya1992experimental}
Nagamiya, S. 1992, Nuclear Physics. A, 544, 5c

\bibitem[{Ostapchenko(2011)}]{ostapchenko2011monte}
Ostapchenko, S. 2011, \prd, 83, 014018, \dodoi{10.1103/PhysRevD.83.014018}

\bibitem[{Patrignani {et~al.}(2016)Patrignani, Agashe, Aielli, Amsler,
  Antonelli, Asner, Baer, Banerjee, Barnett, Basaglia,
  {et~al.}}]{patrignani2016review}
Patrignani, C., Agashe, K., Aielli, G., {et~al.} 2016, Chinese Physics C, 40,
  100001, \dodoi{10.1088/1674-1137/40/10/100001}

\bibitem[{Pierog {et~al.}(2015)Pierog, Karpenko, Katzy, Yatsenko, \&
  Werner}]{pierog2015epos}
Pierog, T., Karpenko, I., Katzy, J.~M., Yatsenko, E., \& Werner, K. 2015, \prc,
  92, 034906, \dodoi{10.1103/PhysRevC.92.034906}

\bibitem[{Rafelski(2016)}]{rafelski2016melting}
Rafelski, J. 2016, {LaTeX: Melting hadrons, boiling quarks: from Hagedorn
  temperature to ultra-relativistic heavy-ion collisions at CERN: with a
  tribute to Rolf Hagedorn}, 1st edn. (Springer Nature)

\bibitem[{Rafelski \& M{\"u}ller(1982)}]{rafelski1982strangeness}
Rafelski, J., \& M{\"u}ller, B. 1982, \prl, 48, 1066,
  \dodoi{10.1103/PhysRevLett.48.1066}

\bibitem[{Roberts(2020)}]{roberts2020insights}
Roberts, C.~D. 2020, Journal of Physics: Conference Series, 1643, 012194,
  \dodoi{10.1088/1742-6596/1643/1/012194}

\bibitem[{Rodr\'\i{}guez-Quintero {et~al.}(2020)Rodr\'\i{}guez-Quintero, Chang,
  Raya, \& Roberts}]{rodriguez2020process}
Rodr\'\i{}guez-Quintero, J., Chang, L., Raya, K., \& Roberts, C.~D. 2020, J.
  Phys. Conf. Ser., 1643, 012177, \dodoi{10.1088/1742-6596/1643/1/012177}

\bibitem[{Sciutto(1999)}]{sciutto1999aires}
Sciutto, S. 1999, arXiv e-prints, \dodoi{10.13140/RG.2.2.12566.40002}

\bibitem[{Thunman {et~al.}(1996)Thunman, Ingelman, \&
  Gondolo}]{thunman1996charm}
Thunman, M., Ingelman, G., \& Gondolo, P. 1996, Astroparticle Physics, 5, 309,
  \dodoi{10.1016/0927-6505(96)00033-3}

\bibitem[{Ulrich {et~al.}(2011)Ulrich, Engel, \& Unger}]{ulrich2011hadronic}
Ulrich, R., Engel, R., \& Unger, M. 2011, \prd, 83, 054026,
  \dodoi{10.1103/PhysRevD.83.054026}

\bibitem[{Van~Hecke(1991)}]{van1991na34}
Van~Hecke, H. 1991, Nucl. Phys. A, 525, 227

\bibitem[{Wong(1994)}]{wong1994introduction}
Wong, C.-Y. 1994, {LaTeX: Introduction to high-energy heavy-ion collisions},
  1st edn. (World Scientific)

\bibitem[{Zajc {et~al.}(1992)}]{E802:1992wow}
Zajc, W.~A., {et~al.} 1992, Nucl. Phys. A, 544, 237,
  \dodoi{10.1016/0375-9474(92)90577-7}

\bibitem[{Zhu {et~al.}(2022)Zhu, Chen, Cui, \& Ruan}]{zhu2022gluon}
Zhu, W., Chen, Q., Cui, Z., \& Ruan, J. 2022, Nucl. Phys. B, 984, 115961,
  \dodoi{10.1016/j.nuclphysb.2022.115961}

\bibitem[{Zhu \& Lan(2017)}]{zhu2017gluon}
Zhu, W., \& Lan, J. 2017, Nucl. Phys. B, 916, 647,
  \dodoi{10.1016/j.nuclphysb.2017.01.021}

\bibitem[{Zhu \& Ruan(1999)}]{zhu1999new}
Zhu, W., \& Ruan, J. 1999, Nuclear Physics B, 559, 378,
  \dodoi{10.1016/S0550-3213(99)00461-7}

\bibitem[{Zhu {et~al.}(2008)Zhu, Shen, \& Ruan}]{wei2008can}
Zhu, W., Shen, Z., \& Ruan, J. 2008, Chin. Phys. Lett., 25, 3605,
  \dodoi{10.1088/0256-307X/25/10/023}

\bibitem[{Zhu {et~al.}(2016)Zhu, Shen, \& Ruan}]{zhu2016chaotic}
---. 2016, Nucl. Phys. B, 911, 1, \dodoi{10.1016/j.nuclphysb.2016.06.031}

\bibitem[{Zhu \& Shen(2004)}]{zhu2004properties}
Zhu, W., \& Shen, Z.-q. 2004, arXiv e-prints, 29, 109,
  \dodoi{10.48550/arXiv.hep-ph/0406213}

\end{thebibliography}



\end{document}